\documentclass[twocolumn,pra,showpacs, twocolumn]{revtex4-1}
\usepackage[T1]{fontenc}
\usepackage[latin9]{inputenc}
\usepackage{amsmath}
\usepackage{amssymb}
\usepackage{graphicx}
\usepackage{esint}

\makeatletter
\@ifundefined{textcolor}{}
{%
 \definecolor{BLACK}{gray}{0}
 \definecolor{WHITE}{gray}{1}
 \definecolor{RED}{rgb}{1,0,0}
 \definecolor{GREEN}{rgb}{0,1,0}
 \definecolor{BLUE}{rgb}{0,0,1}
 \definecolor{CYAN}{cmyk}{1,0,0,0}
 \definecolor{MAGENTA}{cmyk}{0,1,0,0}
 \definecolor{YELLOW}{cmyk}{0,0,1,0}
 }

\makeatother

\begin{document}

\title{EIT-like phenomenon with two atomic ensembles in a cavity}

\author{Yusuf Turek$^{1}$, Yong Li$^{2}$,}

\email{liyong@csrc.ac.cn}

\author{C. P. Sun$^{1,2}$}

\email{cpsun@csrc.ac.cn}

\homepage{http://www.csrc.ac.cn/~suncp}

\affiliation{$^{1}$State Key Laboratory of Theoretical Physics, Institute of
Theoretical Physics, Chinese Academy of Sciences, and University of
the Chinese Academy of Sciences, Beijing 100190, China\\
 $^{2}$Beijing Computational Science Research Center, Beijing 100084,
China}
\begin{abstract}
We study the spectra of collective low excitations of two atomic ensembles
coupled indirectly through a single-mode cavity field. When the left
ensemble is driven with an external optical field, its corresponding
response spectrum to the incident optical light %(reflectivity spectrum)
shows an electromagnetically induced transparency- (EIT-) like phenomenon
when the layers are arranged in the sequence of node-antinode but
not in the sequence of antinode-node. %with the window width controlled by the strength of cavity-ensemble coupling \textbf{(LY note: have we mentioned this in the following?)}.This is an electromagnetically induced transparency- (EIT-) like phenomenon with an asymmetric structure: When the positions of ensembles are exchanged to be in the antinode-node sequence, the EIT-like phenomenon disappear.
In the case of antinode-antinode sequence, the response spectrum shows
an EIT-like phenomenon with two transparent windows. We also investigate
the fluctuation spectra of the atomic collective excitation modes,
which show similar EIT-like phenomena. %These discoveries qualitatively fit the result of recent experiment about the X-ray scattering by nuclei in a cavity (R. R$\ddot{\mathrm{o}}$hlsberger \emph{et al.}, Nature \textbf{482}, 199 (2012)).
%\textbf{(LY note: Can we say it qualitatively fits the result of experiment?)}.

\end{abstract}


\date{\today}

\maketitle

\section{Introduction}

It is well known that photon system is a prior candidate for quantum
information processing such as quantum computing or quantum cryptography
due to its fast and easy-getting advantages. Since the direct coupling
between photons is absent according to the theory of quantum electrodynamics
(QED), people proposed to store the information of photons into an
atomic-ensemble-based quantum memory so that one could indirectly
manipulate photon by photon through the atomic ensemble based on electromagnetically
induced transparency (EIT). Therein the EIT is used to effectively
overcome the strong absorption of an atomic medium on a propagating
beam of electromagnetic radiation~\cite{Harris1997,Fleihhaouer 2005}.
It plays a role in manipulating the photons for quantum information
storage and slow light phenomena~\cite{Lukin 2003,Lukin2000,Mewes 2002,Flei 2001}.
Actually, the EIT phenomenon is a result of the Fano interference
between transitions of atomic internal energy states~\cite{Fano 1961,Lukin 2001na},
and induces many strange optical phenomena in the dispersion medium~\cite{Scully1991,Scully1992}.

The conventional EIT phenomena were implemented for the three-level
or four-level systems which look {}``dark'' for the probe light~\cite{Fleihhaouer 2000,Fleihhaouer 2002}.
Recently the EIT analog in quantum optomechanical systems, i.e., optomechanically
induced transpency (OMIT), was suggested~\cite{Huang} and confirmed
experimentally~\cite{Weis}. Most recently, It was also discovered
that there may also exist the EIT-like phenomena for the reflectivity
spectrum of X-ray in the system of two layers~\cite{Ralf2012} inside
a cavity. Here, the two layers consist of the M$\ddot{\mathrm{o}}$ssbauer
isotope $^{57}$Fe nuclei, which are exactly modeled as two-level
systems with a resonant transition of $14.4$ kev for M$\ddot{\mathrm{o}}$ssbauer
effect. It was observed the EIT-like phenomenon appears for the cavity
configuration where the layers are arranged in the sequence of node-antinode
and disappears in the antinode-node sequence. Such observation was
explained with a proposed three-level configuration, which was usually
required for obtaining EIT effect, but actually the $^{57}$Fe nuclei
for the considered problems are only modeled as two-level systems
rather than three-level ones.

On the other hand, the EIT effect may have a classical analogue that
is referred to three-level configuration directly: A system of two
coupled harmonic oscillators (HOs) can exhibit the EIT-like effect~\cite{Garrido 2002}
wherein a transparency window exists as the coupling induces split
in absorption spectrum. Furthermore, two of the authors of the present
paper (Sun and Li)~\cite{Sun-PRL,Li-Wang-Sun-EPJD} even used the
coupled bosonic modes to describe the low excitations of the atomic
ensemble with EIT configuration. These studies actually gave a description
of two coupled HOs for the EIT effect with atomic ensembles. With
these considerations, it is possible to qualitatively understand the
experiment~\cite{Ralf2012} as an EIT-like phenomenon, that is, as
a classical analogue with coupled bosonic modes formed by the single-mode
cavity field and collective excitation modes of two ensembles/layers
of two-level systems rather than three-level configuration.

In this paper, we consider a model system similar to the two-layer
system in the X-ray quantum optics experiment\cite{Ralf2012}. But
in constrast to the latter case, in our model the two HOs, which are
realized by atomic collective excitation modes, are indirectly coupled
by a quantized single-mode cavity field. We first bosonize the low
excitations of two atomic ensembles inside the single mode cavity.
The quantized single-mode cavity field provides a coupling between
these two bosonic modes. From the quantun Langevin equations of our
system's variables for the two bosonic modes (with one of them driven
by external field) and cavity mode, we find that the steady-state
response intensities of two atomic ensembles show a transparency window
in some conditions, e.g., when the layers are arranged in the sequence
of node-antinode (instead of the sequence of antinode-node).

Indeed, the response spectra of two collective excitation modes to
the external driving show that if the two atomic ensembles are placed
in proper positions, e.g., in the antinode-node sequence, the EIT-like
window would appear for the driven ensemble, and can be explained
as a classical analogue of EIT for two coupled HOs in Ref.\cite{Garrido 2002}.
In the exchanged node-antinode sequence, the EIT-like window disappears.
These could reflect qualitatively the basic spirit hidden in the X-ray
scattering experiment in Ref.\cite{Ralf2012}, where the reflectivity
spectrum shows EIT phenomenon in certain sequence and not in the opposite
sequence. If we put both the atomic ensembles to the antinodes of
the cavity field, the spectra of both the ensembles would appear with
two EIT-like windows. This is very similar to the AC Stark effect
in atomic optics\cite{Fox}. To confirm the the above predictions
based on a simple model, we also calculate the fluctuation spectra,
which display the similar EIT-like phenomena.

This paper is organized as follows. In Sec.~II, we describe our model
with an effective Hamiltonian in terms of the collective excitation
operators of atomic ensembles. In Sec.~III, we calculate the response
spectra of the atomic collective excitation modes to the external
driving field to show the EIT-like phenomena occurred in our scheme.
In Sec.~IV, we calculate the fluctuation spectra of system to confirm
the results in Sec.~III. %We assume that the first (left) and second (right) atomic ensemble's low excitation spectrum represents the absorption and transmission spectrum to incident probe field, respectively.
Finally, we make conclusions and give some remarks in Sec.~V.

\section{The model and Hamiltonian of two ensembles in a cavity}

\begin{figure}
\includegraphics[width=8cm]{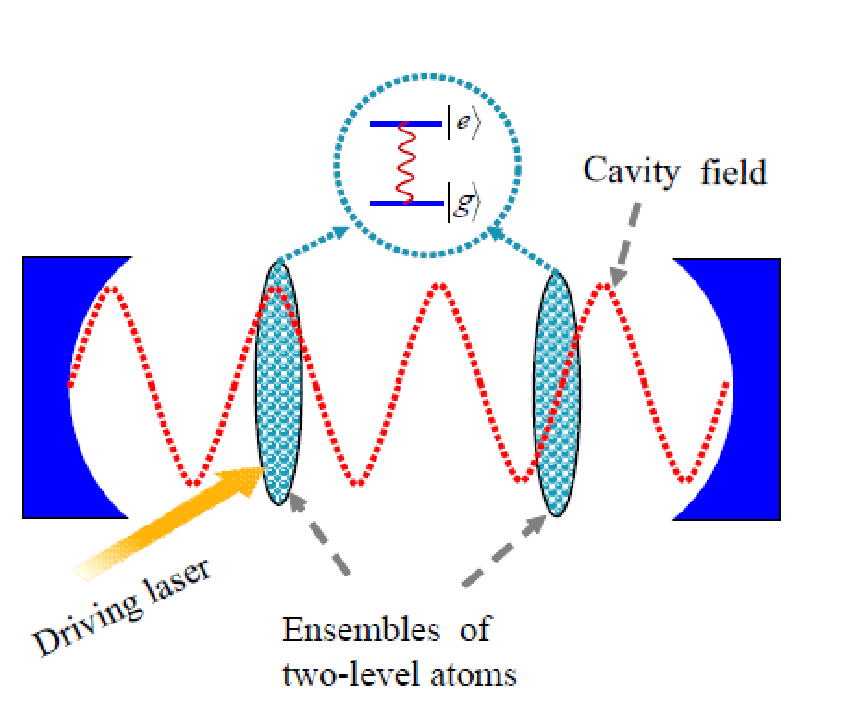} \caption{(Color online) Schematic of two atomic ensembles coupling with single-mode
cavity field: Two atomic ensembles consisted of two-level atoms are
placed in proper positions of single mode cavity field. A classical
radiation field with frequency $\omega_{f}$ drives the left atomic
ensemble.}

\label{model}
\end{figure}

As shown in Fig.~\ref{model}, the model under consideration consists
of two ensembles of two-level atoms coupling with a single-mode cavity
field. The left atomic ensemble is driven by a classical external
field of frequency $\omega_{f}$. The model Hamiltonian reads (hereafter,
we take $\hbar=1$)
\begin{eqnarray}
H & = & \omega_{c}c^{\dagger}c+\frac{\omega_{a}}{2}\sum_{i=1}^{N_{a}}\sigma_{z,a}^{\left(i\right)}+\frac{\omega_{b}}{2}\sum_{j=1}^{N_{b}}\sigma_{z,b}^{\left(j\right)}+\left[g_{a}c\sum_{i=1}^{N_{a}}\sigma_{+,a}^{\left(i\right)}\right.\notag\\
 &  & \left.+g_{b}c\sum_{j=1}^{N_{b}}\sigma_{+,b}^{\left(j\right)}+\Omega e^{-i\omega_{f}t}\sum_{i=1}^{N_{a}}\sigma_{+,a}^{\left(i\right)}+\text{H.c.}\right].\label{eq:Total Hamil}
\end{eqnarray}
Here, $c$ ($c^{\dagger}$) is the annihilation (creation) operator
of the cavity field of frequency $\omega_{c}$. And $\sigma_{z,s}^{\left(l\right)}=\left\vert e\right\rangle _{s}^{(l)}\left\langle e\right\vert -\left\vert g\right\rangle _{s}^{(l)}\left\langle g\right\vert $,
$\sigma_{+,s}^{\left(l\right)}=\left\vert e\right\rangle _{s}^{(l)}\left\langle g\right\vert $,
and $\sigma_{-,s}^{\left(l\right)}=\left\vert g\right\rangle _{s}^{(l)}\left\langle e\right\vert $
($s=a,b$) are the Pauli matrices for the $l$-th atom in the left
ensemble ($s=a$) or the right ensemble ($s=b$) with the same energy
level spacing $\omega_{a}=\omega_{b}$, the number of atoms $N_{a}$/$N_{b}$,
and the excited and ground states of atoms $\left\vert e\right\rangle _{a}$/$\left\vert g\right\rangle _{a}$,
respectively. It is pointed out that each ensemble is arranged in
a thin layer whose size in the direction of cavity axis has been assumed
to be much smaller than the wavelength of cavity field. Thus all the
atoms in the left/right ensemble couple to the single-mode cavity
field with the identical coupling strength $g_{a}$/$g_{b}$. Due
to the same reason, the coupling coefficient between the external
driving field and the atoms in the left ensemble, $\Omega$, is also
identical.

%%%%%%%%%%%%%   Fig. 1 %%%%%%%%%%%%%%%%%%
%\begin{verse}

%\end{verse}

In order to simplify the model Hamiltonian we introduce the following
operators of atomic collective excitation modes~\cite{Sun-PRL,Liu-PRA,Jin_PRB}
for the two atomic ensembles
\begin{equation}
A^{\dagger}=\frac{1}{\sqrt{N_{a}}}\sum_{i=1}^{N_{a}}\sigma_{+,a}^{\left(i\right)},\text{ }A=(A^{\dagger})^{\dagger}\text{,}\label{eq:A boson}
\end{equation}
and
\begin{equation}
B^{\dagger}=\frac{1}{\sqrt{N_{b}}}\sum_{j=1}^{N_{b}}\sigma_{+,b}^{\left(j\right)},\text{ }B=(B^{\dagger})^{\dagger}\text{.}\label{eq:B boson}
\end{equation}
In the low-excitation limit with large $N_{a}$ and $N_{b}$, the
above operators satisfy the standard bosonic commutation relations
\begin{equation}
\left[A,A^{\dagger}\right]\approx\left[B,B^{\dagger}\right]\approx1,\text{ }\left[A,B\right]=\left[A,B^{\dagger}\right]=0.\label{eq:commu rel}
\end{equation}
And we can also have
\begin{eqnarray}
\sum_{i=1}^{N_{a}}\sigma_{z,a}^{\left(i\right)} & = & 2A^{\dagger}A-N_{a},\text{ }\\
\sum_{j=1}^{N_{b}}\sigma_{z,b}^{\left(j\right)} & = & 2B^{\dagger}B-N_{b}.
\end{eqnarray}

Then, Hamiltonian (\ref{eq:Total Hamil}) can be rewritten in terms
of the atomic collective operators $A$ $\left(A^{\dagger}\right)$
and $B$ $\left(B^{\dagger}\right)$ as
\begin{eqnarray}
H & = & \omega_{c}c^{\dagger}c+\omega_{a}A^{\dagger}A+\omega_{b}B^{\dagger}B\notag\\
 &  & +\left(G_{A}cA^{\dagger}+G_{B}cB^{\dagger}+\chi A^{\dagger}e^{-i\omega_{f}t}+\text{H.c.}\right)\label{eq:effec H}
\end{eqnarray}
with $G_{A}\equiv\sqrt{N_{a}}g_{a}$, $G_{B}\equiv\sqrt{N_{b}}$$g_{b}$
and $\chi\equiv\sqrt{N_{a}}\Omega$. For simplicity here we have assumed
all these coupling strengths are real. %The total number of excitation $N=a^{\dagger }a+A^{\dagger }A+B^{\dagger }B$ in the atom-field system is not constant during the evolution of this system, due to external driving field , i.e., $\left[ N,H\right] \neq 0$.
In the interaction picture with respect to $H_{0}=\omega_{f}\left(c^{\dagger}c+A^{\dagger}A+B^{\dagger}B\right)$,
the interaction Hamiltonian is given in the time-independent form
as
\begin{eqnarray}
H_{I} & = & \Delta_{c}c^{\dagger}c+\Delta_{a}A^{\dagger}A+\Delta_{b}B^{\dagger}B\notag\\
 &  & +\left(G_{A}cA^{\dagger}+G_{B}cB^{\dagger}+\chi A^{\dagger}+H.c.\right),\label{eq:Effect H}
\end{eqnarray}
where the detunings $\Delta_{r}\equiv\omega_{r}-\omega_{f}$ for $r=a,b,c$.
We note that during the derivation of Eq. (\ref{eq:effec H}) we have
neglected the constant terms $-\left(1/2\right)\omega_{a}N_{a}$ and
$-\left(1/2\right)\omega_{a}N_{b}$ since they have not any affect
to our result in the context.

\section{The response spectrum for two atomic ensembles in a cavity}

In our model the external optical driving field can be considered
as a probe one which is incident from the left side and drives the
first (left) atomic ensemble. Let us first study the response spectra
of atomic collective excitation modes to the driving. To this end,
we investigate the steady-state solution of variables resorting to
the quantum Langevin equations from Eq.~(\ref{eq:Effect H})
\begin{align}
\overset{\cdot}{c} & =-i\Delta_{c}c-iG_{A}A-iG_{B}B-\frac{\kappa}{2}c+\sqrt{\kappa}c_{\mathrm{in}}(t),\label{eq:Langevin Eq. for c}\\
\overset{\cdot}{A} & =-i\Delta_{a}A-iG_{A}c-i\chi-\frac{\gamma_{A}}{2}A+\sqrt{\gamma_{A}}A_{\mathrm{in}}(t),\label{eq:Langevin Eq. for A-1}\\
\dot{B} & =-i\Delta_{b}B-iG_{B}c-\frac{\gamma_{B}}{2}B+\sqrt{\gamma_{B}}B_{\mathrm{in}}(t).\label{eq:Langiven Eq. for B-1}
\end{align}
Here $\kappa$ is the decay rate of the cavity and $\gamma_{A,B}$
the decay rates of collective modes $A$ and $B$, the operators $c_{\mathrm{in}}\left(t\right)$,
$A_{\mathrm{in}}\left(t\right)$ and $B_{\mathrm{in}}\left(t\right)$
denote the corresponding noises with the vanishing average values,
i.e., $\langle c_{\mathrm{in}}\rangle=$ $\langle A_{\mathrm{in}}\rangle$
$=\langle B_{\mathrm{in}}\rangle=0$. These noise operators satisfy
the following fluctuation relations \begin{subequations}
\begin{eqnarray}
\langle c_{\mathrm{in}}(t)c_{\mathrm{in}}^{\dagger}(t^{\prime})\rangle & = & [N(\omega_{c})+1]\delta(t-t^{\prime}),\\
\langle A_{\mathrm{in}}(t)A_{\mathrm{in}}^{\dagger}(t^{\prime})\rangle & = & [N(\omega_{a})+1]\delta(t-t^{\prime}),\\
\langle B_{\mathrm{in}}(t)B_{\mathrm{in}}^{\dagger}(t^{\prime})\rangle & = & [N(\omega_{b})+1]\delta(t-t^{\prime}),
\end{eqnarray}
where \end{subequations}
\begin{equation}
N(\omega_{r})=\frac{1}{\exp\left(\frac{\omega_{r}}{k_{B}T}\right)-1},\ \ (r=a,b,c)
\end{equation}
are, respectively, the average thermal excitation numbers of the cavity
mode and atomic collective modes at temperature $T$.

The steady-state values of the atomic ensembles-cavity system are
given by
\begin{eqnarray}
A{}_{s}\equiv\langle A\rangle & = & -\frac{\chi F_{A}}{\Delta_{a}-i\frac{\gamma_{A}}{2}},\label{eq:SS for A'}\\
B{}_{s}\equiv\langle B\rangle & = & \frac{\chi f_{a}f_{b}}{\Delta_{\mathrm{\text{\textrm{eff}}}}^{\left(0\right)}-i\frac{1}{2}\kappa_{\text{\textrm{eff}}}^{\left(0\right)}},\label{eq:SS for B'}\\
c{}_{s}\equiv\langle c\rangle & = & \frac{\chi f_{a}}{\Delta_{\mathrm{\text{ \textrm{eff}}}}^{\left(0\right)}-i\frac{1}{2}\kappa_{\text{\textrm{eff}}}^{\left(0\right)}},\label{eq:SS for c'}
\end{eqnarray}
where
\begin{equation}
F_{A}=1+\frac{G_{A}f_{a}}{\Delta_{\mathrm{\text{\textrm{eff}}}}^{\left(0\right)}-i\frac{1}{2}\kappa_{\text{\textrm{eff}}}^{\left(0\right)}}
\end{equation}
is the modified factor of the coupling coefficient between the left
atomic ensemble and external driving field, and
\begin{equation}
f_{a}=\frac{G_{A}}{\Delta_{a}-i\frac{1}{2}\gamma_{A}},\text{ }f_{b}=\frac{G_{B}}{\Delta_{b}-i\frac{1}{2}\gamma_{B}}.
\end{equation}
Here the effective decay rate and detuning between the cavity and
external driving field are given by
\begin{equation}
\kappa_{\text{\textrm{eff}}}^{\left(0\right)}=\kappa+\frac{G_{A}^{2}\gamma_{A}}{\Delta_{a}^{2}+\frac{1}{4}\gamma_{A}^{2}}+\frac{G_{B}^{2}\gamma_{B}}{\Delta_{b}^{2}+\frac{1}{4}\gamma_{B}^{2}},
\end{equation}
and
\begin{equation}
\Delta_{\mathrm{\text{\textrm{eff}}}}^{\left(0\right)}=\Delta_{c}-\frac{G_{A}^{2}\Delta_{a}}{\Delta_{a}^{2}+\frac{1}{4}\gamma_{A}^{2}}-\frac{G_{B}^{2}\Delta_{b}}{\Delta_{b}^{2}+\frac{1}{4}\gamma_{B}^{2}},
\end{equation}
respectively.

Seen from Eqs.~(\ref{eq:SS for A'}), (\ref{eq:SS for B'}), and
(\ref{eq:SS for c'}), the steady-state values of all the three bosonic
modes are proportional to the driving strength of the external probe
field $\chi$. In what follows in this section, we will investigate
the steady-state response spectra (mean excitation populations $|A_{s}|^{2}$
and $|B_{s}|^{2}$) of the two collective excitation modes of atomic
ensembles.

%%%%%%%%%%%%%%  Fig. 2 %%%%%%%%%%%
%\includegraphics[width=4cm]{Steady_state_case_for_two_atomic_ensembles_1}

\begin{figure}
\includegraphics[width=4cm]{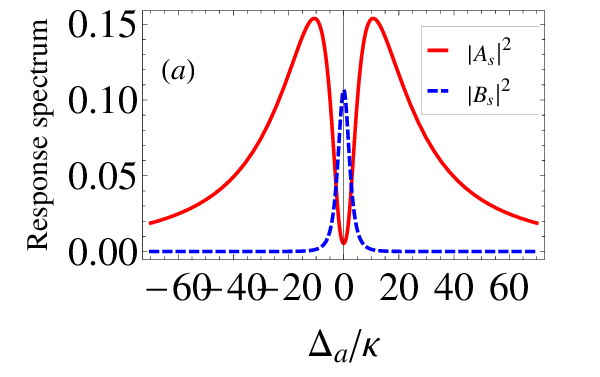} \includegraphics[width=4cm]{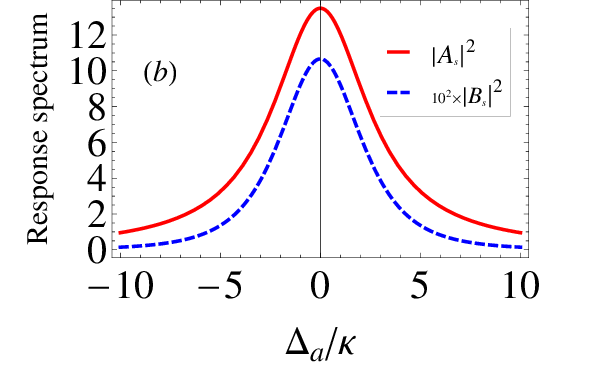}
\caption{(Color online) Plot of the response intensities of atomic ensembles
$|A_{s}|^{2}$ (red solid curve) and $|B_{s}|^{2}$ (blue dashed curve)
in arbitrary units according to Eq.~(\ref{eq:SS for A'}) and Eq.~(\ref{eq:SS for B'})
vs the detuning $\Delta_{a}$. Here, the atomic ensembles are arranged
in (a) the antinode-node sequence with $G_{A}=10$, $G_{B}=1$, $\gamma_{A}=90$,
and $\gamma_{B}=9$, or in (b) the node-antinode sequence with $G_{A}=1$,
$G_{B}=10$, $\gamma_{A}=9$, $\gamma_{B}=90$. All the frequencies
are in units of $\kappa$. And we assumed the degenerated case $\omega_{a}=\omega_{b}=\omega_{c}=10^{7}$,
that is, $\Delta_{a}=\Delta_{b}=\Delta_{c}$.}

\label{Fig2}
\end{figure}

%%%%%%%%%%%%  Fig. 3 %%%%%%%%%%%%%%
%\begin{figure}[tbp]
%\includegraphics[width=8cm]{Steady_state_case_for_two_atomic_ensembles_3}
%\caption{(Color online) Similar to Fig.~\ref{fig:response large A}
%except $G_{A}=1$, $G_{B}=10$, $\gamma_{A}=9$, $\gamma_{B}=90$. This %corresponds to the case of exchanging the positions of two atomic
%ensembles.}
%\label{fig:response large B}
%\end{figure}

Let us first consider the case of antinode-nodesequence that the left
atomic ensemble is placed at an antinode and the right one close to
a node of the single-mode cavity field. That is, the left (right)
ensemble is strongly (weakly) coupled to the cavity field. Seen from
Fig.~\ref{Fig2}(a), in this case the response of the left ensemble
appears with an EIT window, which is similar to the case of two coupled
HOs in Ref.~\cite{Garrido 2002} although our system behaves as a
system of three bosonic modes. This is expected since the present
system will reduce to the model of two coupled bosonic modes when
the right ensemble is placed close to the node with a very small coupling
to the cavity and leads to negligible contribution. As discussed in
Ref.\cite{Garrido 2002},\textbf{ }this effect is similar to the AC
Stark splitting in quantum optics~\cite{Fox}.

Note that the above EIT-like phenomenon for the left driven ensemble
happens when its coupling strength to the cavity field is larger than
the decay rate of the field: $G_{A}>\kappa$, even when $\gamma_{A}>\kappa,G_{A}$.
If we exchange the positions of the two atomic ensemble to make it
in the node-antinode sequence so that $G_{A}\lesssim\kappa,\gamma_{A}$,
the EIT-like phenomenon for the left ensemble would disappear, as
in Fig.~\ref{Fig2}(b). This phenomenon results from the weak coupling
of the driven ensemble to the cavity compared with the corresponding
decay rates. It is noted that in both the above cases the response
of the right ensemble is still a Lorentz-type peak without EIT window
due to the weak coupling of the right ensemble to the cavity field
(see the blue dash line in Fig.~\ref{Fig2}(a)) or the weak coupling
of the left driven ensemble to the cavity field (see the blue dash
line in Fig.~\ref{Fig2}(b)).

\begin{figure}
\includegraphics[width=4cm]{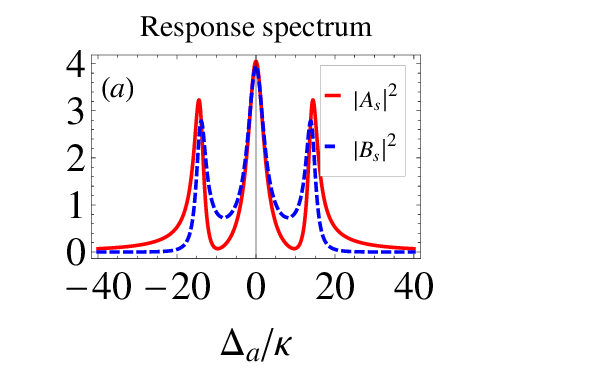} \includegraphics[width=4cm]{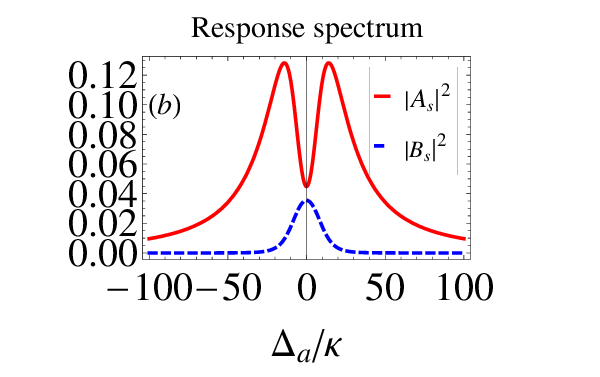}
\includegraphics[width=4cm]{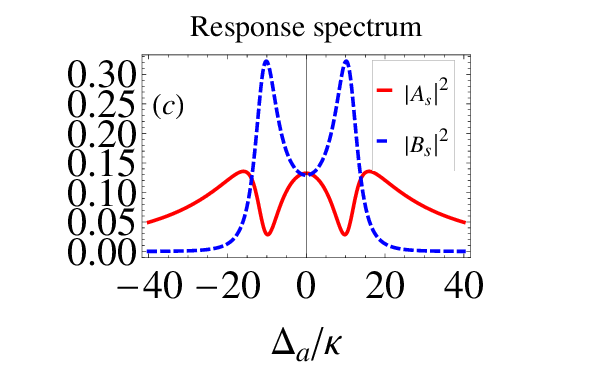} \includegraphics[width=4cm]{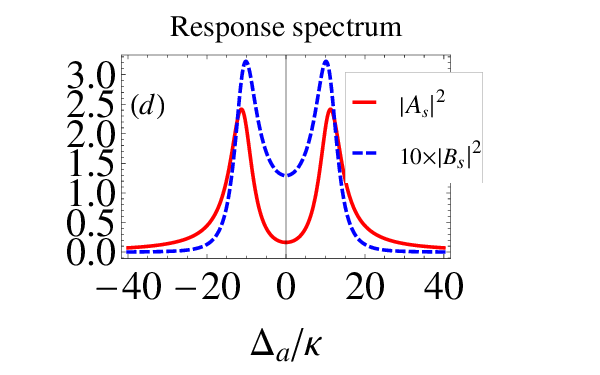}
\caption{(Color online) %Plot of the response intensities of atomic ensembles $|A_{s}^{2}|$ (red solid curve) and $|B_{s}^{2}|$ (blue dashed curve) in arbitrary units. Here, the atomic ensembles are arranged in the antinode-antinode sequence with $G_{A}=G_{B}=10$ but with different decays rates: (a) $\gamma_{A}=5$, and $\gamma_{B}=5$; (b) $\gamma_{A}=50$, $\gamma_{B}=50$; (c) $\gamma_{A}=50$, $\gamma_{B}=5$; (d) $\gamma_{A}=5$, $\gamma_{B}=50$. All the frequencies are in units of $\kappa$. And we assumed the degenerated case $\omega_{a}=\omega_{b}=\omega_{c}=10^{7}$, that is, $\Delta_{a}=\Delta_{b}=\Delta_{c}$.
Similar to Fig.~\ref{Fig2} except for the atomic ensembles arranged
in the antinode-antinode sequence with $G_{A}=G_{B}=10$ but with
different decays rates: (a) $\gamma_{A}=5$, and $\gamma_{B}=5$;
(b) $\gamma_{A}=50$, $\gamma_{B}=50$; (c) $\gamma_{A}=50$, $\gamma_{B}=5$;
(d) $\gamma_{A}=5$, $\gamma_{B}=50$.}

\label{Fig3}
\end{figure}

Now let us consider the response spectra of the ensembles when both
the atomic ensembles position at (or near) the antinodes of the cavity
field, e.g., $G_{A}=G_{B}=10\kappa$ in Fig.~\ref{Fig3}. If both
the decay rates of the atomic collective excitation modes are not
so large, e.g, $\gamma_{A}=\gamma_{B}=5\kappa$ ($<G_{A}$, $G_{B}$)
as shown in Fig.~\ref{Fig3}(a), the response spectra of both the
ensembles appear with two pronounced EIT-like windows which are expected
to occur in a system of three coupled HOs. This is a simple generalization
of the classical analog of EIT-like mechanism in a system of two coupled
HOs~\cite{Garrido 2002}. However, when the decay rates of the ensembles
are very large, e.g, $\gamma_{A}=\gamma_{B}=50\kappa$ ($>G_{A}$,
$G_{B}$) in Fig.~\ref{Fig3}(b), the response of the driven ensemble
appears with only one EIT-like window and the one of the right ensemble
appears without any EIT window. In the case of $\gamma_{A}=50\kappa$
and $\gamma_{B}=5\kappa$ in Fig.~\ref{Fig3}(c), the driven ensemble
has the response with two-window EIT-like phenomenon and the right
one with one-window EIT-like phenomenon. Shown in Fig.~\ref{Fig3}(d)
with $\gamma_{A}=5\kappa$ and $\gamma_{B}=50\kappa$, due to the
fact that the strong decay rate of the right ensemble destroys its
action on the other modes, the response of the driven ensemble happens
with one EIT-like window, similar to the case of two coupled HOs.
At the same time, the strong coupling of the right ensemble to the
cavity field makes it have a weak EIT-like response to the external
field driving on the left ensemble.

%%%%%%%%%%%%%%%  Fig. 3 %%%%%%%%%%%%%%%%%%%
%\includegraphics[width=8cm]{Steady_state_case_for_two_atomic_ensembles_2013_01_04_02}

\section{The fluctuation spectrum for two atomic ensembles in a cavity}

In this section, we will consider the fluctuation spectra of the atomic
ensembles in the cavity when the left ensemble is driven. To account
for the effects of the quantum fluctuations we decompose each bosonic
operator in the Langevin equations~(\ref{eq:Langevin Eq. for c}),~(\ref{eq:Langevin Eq. for A-1}),
and~(\ref{eq:Langiven Eq. for B-1}) as the sum of its steady-state
value and a small fluctuation, e.g., $c=c_{s}+\delta c$. Substituting
these quantities into the Langevin equations and linearizing the resulting
equations for the fluctuations, one has
\begin{eqnarray}
\delta\dot{c}=-i\Delta_{c}\delta c-iG_{A}\delta A-iG_{B}\delta B-\frac{\kappa}{2}\delta c+\sqrt{\kappa}c_{\mathrm{in}}(t),\\
\delta\dot{A}=-i\Delta_{a}\delta A-iG_{A}\delta c-\frac{\gamma_{A}}{2}\delta A+\sqrt{\gamma_{A}}A_{\mathrm{in}}(t),\label{eq:linearized QLA c}\\
\delta\dot{B}=-i\Delta_{b}\delta B-iG_{B}\delta c-\frac{\gamma_{B}}{2}\delta B+\sqrt{\gamma_{B}}B_{\mathrm{in}}(t).\label{eq:linearized QLE b}
\end{eqnarray}

For the experimental perspective the frequency domain is more useful.
Thus by Fourier-transferring these equations into the frequency domain
like
\begin{equation}
\tilde{y}\left(\omega\right)=\frac{1}{2\pi}\int_{-\infty}^{+\infty}y\left(t\right)e^{i\omega t}dt\label{eq:Fourier Trans}
\end{equation}
for any operator $y\left(t\right)$, it is easy to find the following
solutions \begin{widetext}
\begin{eqnarray}
\delta\tilde{c}\left(\omega\right) & = & \frac{i}{\omega-\Delta{}_{\mathrm{eff}}(\omega)+i\frac{1}{2}\kappa_{\text{\textrm{eff}}}(\omega)}\times\left[\sqrt{\kappa}\tilde{c}_{\mathrm{in}}(\omega)+\frac{G_{A}\sqrt{\gamma_{A}}\tilde{A}_{\mathrm{in}}(\omega)}{\left(\omega-\Delta_{a}\right)+i\frac{\gamma_{A}}{2}}+\frac{G_{B}\sqrt{\gamma_{B}}\tilde{B}_{\mathrm{in}}(\omega)}{\left(\omega-\Delta_{b}\right)+i\frac{\gamma_{B}}{2}}\right],\\
\delta\tilde{A}\left(\omega\right) & = & \frac{G_{A}\delta\tilde{c}(\omega)+i\sqrt{\gamma_{A}}\tilde{A}_{\mathrm{in}}(\omega)}{\left(\omega-\Delta_{a}\right)+i\frac{\gamma_{A}}{2}},\ \ \ \ \delta\tilde{B}\left(\omega\right)=\frac{G_{B}\delta\tilde{c}(\omega)+i\sqrt{\gamma_{B}}\tilde{B}_{\mathrm{in}}(\omega)}{\left(\omega-\Delta_{b}\right)+i\frac{\gamma_{B}}{2}},
\end{eqnarray}
%\end{widetext}
where %\begin{widetext}
%\begin{subequations}
\begin{eqnarray}
\Delta_{\text{\textrm{eff}}}(\omega) & = & \Delta_{c}+\frac{G_{A}^{2}\left(\omega-\Delta_{a}\right)}{\left(\omega-\Delta_{a}\right)^{2}+\frac{\gamma_{A}^{2}}{4}}+\frac{G_{B}^{2}\left(\omega-\Delta_{b}\right)}{\left(\omega-\Delta_{b}\right)^{2}+\frac{\gamma_{B}^{2}}{4}},\\
\kappa_{\text{\textrm{eff}}}(\omega) & = & \kappa+\frac{\gamma_{A}G_{A}^{2}}{\left(\omega-\Delta_{a}\right)^{2}+\frac{\gamma_{A}^{2}}{4}}+\frac{\gamma_{B}G_{B}^{2}}{\left(\omega-\Delta_{b}\right)^{2}+\frac{\gamma_{B}^{2}}{4}}.
\end{eqnarray}
%\end{widetext}

Now we calculate the fluctuation spectra of the cavity field and the
atomic collective-excitation modes, $S_{c,A,B}\left(\omega\right)$,
which are defined as \cite{Orszag}
\begin{equation}
S_{y}\left(\omega\right)=\frac{1}{2\pi}\int_{-\infty}^{+\infty}\langle\delta y\left(t-\tau\right)\delta y^{\dagger}\left(t\right)\rangle e^{i\omega\tau}d\tau,\ \ (y=c,A,B).
\end{equation}

The explicit forms of the fluctuation spectra of the collective-excitation
modes for the atomic ensemble are
\begin{eqnarray}
S_{A}(\omega) & = & \frac{G_{A}^{2}S_{c}(\omega)+\gamma_{A}[N(\omega_{a})+1]\left[1+2G_{A}^{2}K_{A}(\omega)\right]}{\left(\omega-\Delta_{a}\right)^{2}+\frac{1}{4}\gamma_{A}^{2}},\label{eq:fluctuata spec of A}\\
S_{B}(\omega) & = & \frac{G_{B}^{2}S_{c}(\omega)+\gamma_{B}[N(\omega_{b})+1]\left[1+2G_{B}^{2}K_{B}(\omega)\right]}{\left(\omega-\Delta_{b}\right)^{2}+\frac{1}{4}\gamma_{B}^{2}},\label{eq:Fluc spec of B}
\end{eqnarray}
where
\begin{equation}
S_{c}(\omega)=\frac{[N(\omega_{c})+1]\kappa+\frac{G_{A}^{2}\gamma_{A}}{\left(\omega-\Delta_{a}\right)^{2}+\frac{\gamma_{A}^{2}}{4}}[N(\omega_{a})+1]+\frac{G_{B}^{2}\gamma_{B}}{\left(\omega-\Delta_{b}\right)^{2}+\frac{\gamma_{B}^{2}}{4}}[N(\omega_{b})+1]}{\left[\omega-\Delta_{\text{\textrm{eff}}}(\omega)\right]^{2}+\frac{1}{4}\kappa_{\text{\textrm{eff}}}^{2}(\omega)},\label{eq:fluc spec of c}
\end{equation}
is the fluctuation spectrum of the cavity field. Here we take \begin{subequations}
\begin{eqnarray}
K_{A}(\omega) & = & \frac{\left(\omega-\Delta_{a}\right)\left[\omega-\Delta_{\text{\textrm{eff}}}(\omega)\right]-\frac{1}{4}\gamma_{A}\kappa_{\text{\textrm{eff}}}(\omega)}{\left(\left[\omega-\Delta_{\text{\textrm{eff}}}(\omega)\right]^{2}+\frac{1}{4}\kappa_{\text{\textrm{eff}}}^{2}(\omega)\right)\left[\left(\omega-\Delta_{a}\right)^{2}+\frac{1}{4}\gamma_{A}^{2}\right]},\\
K_{B}(\omega) & = & \frac{\left(\omega-\Delta_{b}\right)\left[\omega-\Delta_{\text{\textrm{eff}}}(\omega)\right]-\frac{1}{4}\gamma_{B}\kappa_{\text{\textrm{eff}}}(\omega)}{\left(\left[\omega-\Delta_{\text{\textrm{eff}}}(\omega)\right]^{2}+\frac{1}{4}\kappa_{\text{\textrm{eff}}}^{2}(\omega)\right)\left[\left(\omega-\Delta_{b}\right)^{2}+\frac{1}{4}\gamma_{B}^{2}\right]}.
\end{eqnarray}
\end{subequations} \end{widetext}

For simplicity, in this paper we just consider the simple resonant
case of $\omega_{a}=\omega_{b}=\omega_{c}$, that is, $N(\omega_{a})=$
$N(\omega_{b})=$ $N(\omega_{c})$. Note that the thermal photon number
for optical field (like visible light or X-ray) is approximately zero
even at room temperatures. So we just consider this noise response
spectrum with $N(\omega_{a})=$ $N(\omega_{b})=$ $N(\omega_{c})=0$.

\begin{figure}
\includegraphics[width=4cm]{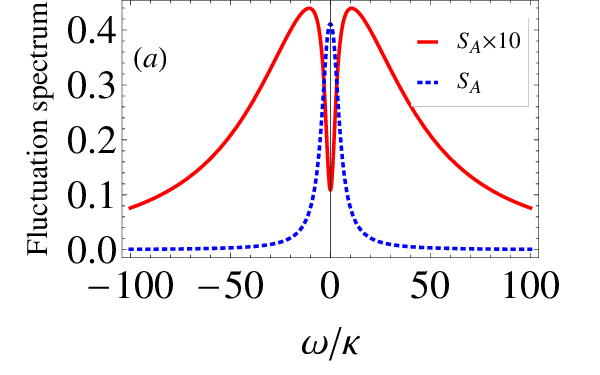}\includegraphics[width=4cm]{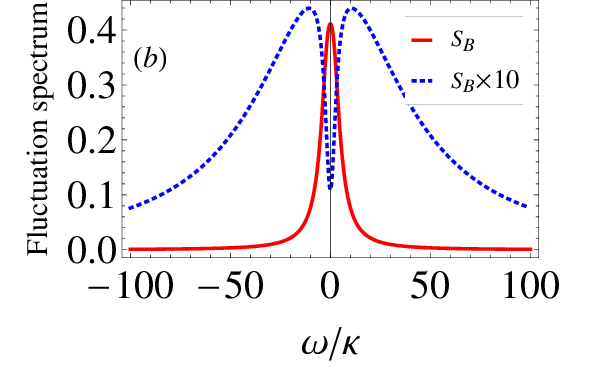}
\caption{(Color online) The fluctuation spectra of the driven ensembles, (a)
$S_{A}(\omega)$ in Eq.~(\ref{eq:fluctuata spec of A}) and (b) $S_{B}(\omega)$
in Eq.~(\ref{eq:Fluc spec of B}) in arbitrary units. Here, $\omega_{c}=10^{7}$,
and $\Delta_{c}=\Delta_{a}=\Delta_{b}=0$ (in units of $\kappa$).
The red solid curve corresponds to the antinode-node sequence with
the parameters $G_{A}=10$, $G_{B}=1$, $\gamma_{A}=90$ and $\gamma_{B}=9$,
and the blue dotted curve corresponds to the node-antinode sequence
with the parameters $G_{A}=1$, $G_{B}=10$, $\gamma_{A}=9$ and $\gamma_{B}=90$.}

\label{fig:the fluc spec of A}
\end{figure}

Seen from Fig.~\ref{fig:the fluc spec of A}(a), the fluctuation
spectrum of the driven left ensemble appears with asymmetrical configuration:
The EIT-like phenomenon appears in the case of antinode-node sequence
(the red solid curve) and disappears in the opposite sequence (the
blue dotted curve), as similar to the response spectra as given in
Fig.~\ref{Fig2}(a). However, the fluctuation spectrum of the right
atomic ensemble in Fig. \ref{fig:the fluc spec of A}(b), happens
with some different features compared to the response spectrum of
the right atomic ensemble as shown in Fig. \ref{Fig2}: The former
will appear with EIT-like window in the case of note-antnode sequence
and the latter will only have Lorentz-type spectrum for the same,
given parameters.

It follows from Eq.(\ref{eq:fluctuata spec of A}) and Eq. (\ref{eq:Fluc spec of B})
that the fluctuation spectra of excitation for the right and the left
atomic ensembles have similar expressions since they equally couple
to single mode cavity field. From Fig. 4 we can see that the spectra
have symmetric two peaks at $\omega/\kappa=\pm G_{A}$ (see the red
solid curve in Fig. \ref{fig:the fluc spec of A}(a)) for the left
atomic ensemble and at $\omega/\kappa=\pm G_{B}$ (see the blue dotted
line in Fig. \ref{fig:the fluc spec of A}(b)) for the right one,
respectively. When we further enhance the coupling coefficients $G_{A}$
and $G_{B}$, the EIT-like windows which occur above would disappear.\textbf{
}Thus the EIT-like windows are controlled by the amount coupling coefficients
under certain conditions.

%%%%%%%%%%%%%  Fig. 4   %%%%%%%%%%%%%%%%

%%%%%%%%%%%%%  Fig. 5   %%%%%%%%%%%%%%%%
%\begin{figure}[tbp]
%\includegraphics[width=8cm]{Two_atomic_ensemble_spectrums_2013_01_31_for_B_atomic_ensemble}
%\caption{(Color online)
%The plot of Eq. $\left(\ref{eq:Fluc spec of B}\right)$ Here, in unit of $\kappa$,$\omega_{c}=10^{5},k_{B}T=2.5\times10^{7},\Delta_{c}=\Delta_{a}=\Delta_{b}=0.$ The red thick curve is corresponding to the parameters $G_{A}=10,G_{B}=1, \gamma_{B}=9,\gamma_{A}=90,$ and the blue thick dotted curve corresponding to the parameters $G_{A}=1,G_{B}=10,\gamma_{B}=90, \gamma_{A}=9$. }
%\label{fig:the fluc spec of B}
%\end{figure}

\section{Conclusion with a remark}

In conclusion, the EIT-like phenomena was shown to happen for two
two-level atomic ensembles inside a cavity when one of them is driven
by a laser. The theoretical prediction was made for realistic systems
without referring an assumption of three-level configuration, which
usually is used for the argument of EIT mechanism. We attribute the
EIT-like phenomena to a simplified model: the three coupled HOs consisting
of the single cavity mode and two collective low-excitation modes
of the two-level atomic ensembles.

Essentially, the EIT (or EIT-like) phenomenon we studied here for
the two ensembles of two-level atoms inherently has the same mechanism
as the the EIT effect for a single ensemble with three-level configuration.
This is because both could be modeled mathematically as the system
of two coupled HOs. In this sence, it is possible to find such EIT-like
phenomena in various hybrid systems, such as an atomic ensemble coupled
to a nano-mechanical resonator or a superconducting transmission line.
\begin{acknowledgments}
This work is supported by National Natural Science Foundation of China
under Grants No.11121403, No. 10935010, No. 11174027, No. 11074261
and National 973 program under Grant No. 2012CB922104.\end{acknowledgments}

\end{document}